# Effectiveness of interactive tutorials in promoting "which-path" information reasoning in advanced quantum mechanics

Alexandru Maries,[1] Ryan Sayer,[2] and Chandralekha Singh[3]

[1]*Department of Physics, University of Cincinnati, Cincinnati, Ohio 45221, USA*
[2]*Department of Physics, Bemidji State University, Bemidji, Minnesota 56601, USA*
[3]*Department of Physics and Astronomy, University of Pittsburgh, Pittsburgh, Pennsylvania 15260, USA*
(Received 24 May 2016; published 11 September 2017)

Research suggests that introductory physics students often have difficulty using a concept in contexts different from the ones in which they learned it without explicit guidance to help them make the connection between the different contexts. We have been investigating advanced students' learning of quantum mechanics concepts and have developed interactive tutorials which strive to help students learn these concepts. Two such tutorials, focused on the Mach-Zehnder interferometer (MZI) and the double-slit experiment (DSE), help students learn how to use the concept of "which-path" information to reason about the presence or absence of interference in these two experiments in different situations. After working on a pretest that asked students to predict interference in the MZI with single photons and polarizers of various orientations placed in one or both paths of the MZI, students worked on the MZI tutorial which, among other things, guided them to reason in terms of which-path information in order to predict interference in similar situations. We investigated the extent to which students were able to use reasoning related to which-path information learned in the MZI tutorial to answer analogous questions on the DSE (before working on the DSE tutorial). After students worked on the DSE pretest they worked on a DSE tutorial in which they learned to use the concept of which-path information to answer questions about interference in the DSE with single particles with mass sent through the two slits and a monochromatic lamp placed between the slits and the screen. We investigated if this additional exposure to the concept of which-path information promoted improved learning and performance on the DSE questions with single photons and polarizers placed after one or both slits. We find evidence that both tutorials promoted which-path information reasoning and helped students use this reasoning appropriately in contexts different from the ones in which they had learned it.

## I. INTRODUCTION

Prior research suggests that in quantum mechanics, students have many common difficulties due the unintuitive and abstract nature of the subject [1–19]. Our group has been investigating advanced students' (upper-level undergraduate and graduate students) reasoning difficulties with quantum mechanics concepts and has been developing and evaluating research-based tools to help students learn quantum mechanics effectively [1,3,4,10–15,20–29]. In particular, we have developed research-validated interactive tutorials, referred to as Quantum Interactive Learning Tutorials, or QuILTs [22,26,27], which strive to help students develop a solid grasp of quantum mechanics concepts. The QuILTs use a guided inquiry-based approach to learning and often include interactive simulations which are pedagogically integrated in the tutorials. Within the guided inquiry-based approach used in the QuILTs, students are asked to make predictions and use the simulations to check their predictions, after which the tutorial provides guidance to help them reconcile any differences between their predictions and observations. In particular, the QuILTs provide students with prompt feedback and support as they strive to repair, extend and organize their knowledge structure related to quantum mechanics. If the QuILTs are effective at helping students develop a good understanding of quantum concepts, it is possible that students would be able to apply the concepts learned from a particular context in the QuILT to different contexts. The extent to which this occurs in the contexts of single photon interference for the double-slit experiment (DSE) [30] and Mach-Zehnder Interferometer (MZI) [31] is the focus of this investigation.

One fundamental concept of quantum mechanics is interference of single particles, e.g., photons, which can be understood using "which-path" information (WPI) reasoning promoted by Wheeler [32] (we provide a definition of WPI and how it can be used to reason about interference in the MZI and DSE in Sec. II). The MZI and DSE provide great contexts to investigate whether the tutorials help advanced students use WPI reasoning that they learned in one context to answer questions in a different context. For example, advanced students learn how to reason in terms of WPI to predict interference in the MZI with single particles and polarizers of various orientations placed in one or both paths of the MZI from an MZI QuILT [33]. We investigated the extent to which students were able to apply similar reasoning to reason about interference in the context of the DSE with single photons and polarizers of various orientations placed after one or both slits. Furthermore, the advanced students learned how to use WPI in a different context of the DSE, namely, particles with mass and a monochromatic lamp placed between the slits and the screen from a DSE QuILT [34]. We investigated the extent to which this additional exposure to WPI in a different context of the DSE helps them make connections with the DSE with single photons and polarizers. More details about the design of this investigation and the QuILTs are provided in Sec. III. Both QuILTs are available in [35]. After following the link, one can request access, which is granted after verifying that the person requesting access is an instructor. Here we describe research that suggests that the QuILTs were effective in helping students learn how to appropriately use WPI reasoning to explain whether interference is observed in a given situation in contexts different from the ones in which it was explicitly learned.

Below, we begin by describing the isomorphism between questions about interference in the MZI and the DSE, after which we discuss the methodology used and research questions investigated. We then describe the findings using quantitative data and qualitative findings from in-depth think-aloud interviews [36], which provide some insight into possible reasons for the effectiveness of the DSE tutorial in promoting WPI reasoning. For those interested, we have included an in-depth analysis of the common student difficulties with the questions posed in this investigation in Appendix C.

## II. ISOMORPHISM BETWEEN MZI AND DSE QUESTIONS

We note that a comprehensive discussion of the isomorphism between questions about interference in the MZI and DSE contexts and how WPI reasoning can be used to predict interference is provided in Appendix A. Below, we provide only a brief explanation.

Before recognizing the isomorphism between questions about interference in the MZI and the DSE contexts, one must first understand how the concept of WPI can be used to reason about interference in each experiment. The concept of WPI at a detector may be useful when the state of the system is a superposition of two different spatial path states (e.g., MZI, DSE with single photons). In general, when a detector can project both components of the path state, then WPI is unknown. On the other hand, when a detector can project only one component of the path state, then we have complete which path information, or WPI is known. For example, when there are no polarizers placed in front of either slit in the DSE or in either path of the MZI, the state of a photon before being detected is a product of a linear superposition of path states with a linear superposition of polarization states (vertical, horizontal). So each polarization state component is associated with both path state components. Thus, the detector or screen can project both path state components for each polarization state, which means that WPI is not known for either polarization. Thus, full interference will be observed on the screen. Suppose that instead, a vertical polarizer is placed in front of one slit (say, upper slit) in the DSE or is placed in one of the paths (say, upper path) of the MZI. Since vertical and horizontal polarization states are orthogonal, placing the vertical polarizer (in front of the upper slit in the DSE or in the upper path of the MZI) will cancel the horizontal component associated with only the upper path state. Thus, for the horizontal component of the photon state, the detector can only project the lower path state component, which means that WPI is known for horizontally polarized photons detected at the screen. On the other hand, WPI is not known for vertically polarized photons because for the vertical component of the photon state, the detector can project both path states. Thus, the horizontally polarized photons detected at the screen will not interfere, whereas the vertically polarized photons will interfere. Similar reasoning can be applied in other situations (see Q1 through Q5 described in Sec. III B).

## III. PARTICIPANTS, MATERIALS, RESEARCH QUESTIONS AND STUDY DESIGN

### A. Participants

The participants in this study were 46 undergraduate students enrolled in an upper-level quantum mechanics course (mostly juniors and seniors in physics) and 59 physics graduate students enrolled in a mandatory semester-long TA professional development course which met for two hours each week. With very few exceptions (several students), all of the students had typical ages you would expect for the level they are at, around 20–22 years for the undergraduate students and 23–25 for the graduate students. For the undergraduate students, the MZI and DSE were part of the course material and therefore the QuILTs and post-tests (described in detail below) were graded for correctness and the post-tests were counted as regular quizzes. In addition, the undergraduate students were aware that topics discussed in these QuILTs can also appear in future exams. After completing a pretest on a particular QuILT, students worked on that tutorial during an hour long class and whatever they did not finish, they completed at home. None of the undergraduate students completed either QuILT in class. For the graduate students, one of the topics of the TA professional development course was the benefits of using the tutorial approach to teaching physics. They were required to engage with the two QuILTs that were on topics with which they were expected to be somewhat familiar but do not fully understand (MZI and DSE) in contrast with engaging with tutorials in introductory physics for which many graduate students are likely to be experts (although there was brief discussion of introductory physics tutorials in class). If graduate students engage with tutorials on topics they do not fully understand, they can learn the topics discussed and understand the value of utilizing these tools as supplements to instruction. For the graduate students, the pretests and post-tests and the QuILTs were graded for completeness instead of correctness since the course performance was graded as satisfactory or unsatisfactory. We note that we refer to questions about interference in the context of the DSE/MZI with single photons and polarizers of various orientations placed in front of one or both slits (DSE) or placed in one or both paths (MZI) as the DSE/MZI polarizer questions (depending on the context in which they were asked). These questions were part of the DSE and MZI pretests and post-tests. Similar to the undergraduate students, the graduate students also completed a pretest at the beginning of a class, after which they worked in groups on a tutorial, which they completed at home and submitted as homework if they did not finish it during class.

## B. Materials

The materials used in this study are two research-based QuILTs on the MZI and DSE, each of which include pretests and post-tests. The DSE pretests and post-tests also include the DSE polarizer questions, a topic which is not discussed in the DSE tutorial (the DSE pretest is included in Appendix B). These polarizer questions (described in detail below) were designed specifically to investigate the extent to which students are able to use WPI reasoning they learned in the context of the MZI to answer isomorphic questions in the context of the DSE. Both the MZI and DSE QuILTs focus on helping students learn about topics such as the wave-particle duality (in the context of single photons in the MZI and in the context of particles with mass in the DSE), interference of single photons (MZI)-particles with mass (DSE), probabilistic nature of quantum measurements, and collapse of a quantum state upon measurement. Both QuILTs make use of interactive simulations in which students can manipulate the MZI and DSE setups to predict and observe what happens at the photo-detectors (MZI)-screen (DSE) for various setups. The development of both QuILTs included interviews with both graduate and undergraduate students in which students worked on the tutorials while thinking out loud. While students worked on the tutorials, they were not disturbed. After they were finished, they were asked for clarification on points they had not made clear earlier while thinking out loud. For the DSE QuILT alone, approximately 85 h of individual student interviews were conducted, each interview lasting 2–4 h. Similar interviews were conducted while developing the MZI QuILT as well. In addition, five physics faculty members were consulted several times during the development of each of these tutorials to ensure that they also found the wording of the questions unambiguous. In addition, their feedback was helpful in ensuring that the topics in the tutorials were addressed appropriately and unambiguously.

In the MZI QuILT, students learn how photodetectors and optical elements such as beam splitters in the path of the MZI with single photons affect measurement outcomes. In addition, the MZI QuILT discusses setups in which polarizers of various orientations are placed in one or both paths and guides students to reason in terms of WPI to predict the outcome at the detectors. Thus, the MZI QuILT provides explicit help for answering the MZI polarizer questions which are isomorphic to the DSE polarizer questions. We hypothesized that if students learn how to reason in terms of WPI to answer the MZI polarizer questions, they may be able to use this reasoning correctly to answer the DSE polarizer questions. Investigation of the extent to which the MZI tutorial may promote use of WPI reasoning was one of the main goals of our investigation (more details on the study design are provided in Sec. III C).

In the DSE QuILT, students learn the basics of single particle interference in the context of the DSE experiment and how different parameters (e.g., mass and kinetic energy of the particles, slit separation, etc.) affect the interference pattern observed on the screen. In addition, they learn how placing a monochromatic lamp between the slits and the screen, which emits photons that scatter with the particles sent through the slits, can alter and in some situations destroy the interference pattern. The reasoning used to help students make sense of the photon-particle scattering in the DSE is also based on WPI. However, the WPI reasoning in the context of the DSE (with particles with mass and a monochromatic lamp placed between the slits and the screen) is for a completely different task and in a very different context than in the context of the MZI with single photons (for more details about the DSE tutorial, see Ref. [34]). It is important to emphasize that in the DSE QuILT, students do not learn about interference for single photons or polarizers in front of slits at all. Thus, the DSE QuILT provides no explicit support for answering the DSE polarizer questions included in the DSE pretest and post-test. However, some students may be able to discern the underlying physics using WPI reasoning (when learning how to apply it to reason about interference in the DSE questions for particles with mass) and apply this reasoning correctly to the DSE polarizer questions. This was the focus of one of our research questions as discussed in Sec. III C.

Each of the MZI and DSE QuILTs includes a pretest and a post-test which includes many questions based on the goals of each tutorial. The DSE pretest and post-test has two major parts (we include the DSE pretest in Appendix B):

(i) Questions related to the impact on the interference pattern of single particles (with mass) due to the addition of a monochromatic lamp close to the slits so that single particles passing through the slits scatter off photons emitted by the light source. We refer to these questions as the "DSE lamp questions." These questions were explicitly discussed in the DSE QuILT, which helped students make sense of them by using reasoning related to WPI.

(ii) DSE polarizer questions related to interference of single photons passing through the slits and the effect on the interference pattern of placing polarizers of various orientations after one or both slits. These topics were not discussed in the DSE QuILT.

The DSE polarizer questions are summarized as follows:

"You perform a DSE in which photons that are polarized at $+45°$ are sent one at a time towards the double slit. The wavelength of the photons is comparable to the slit width and the separation between the slits is more than twice the slit width. In all questions, assume that the same large number $N$ of photons reaches the screen. In each situation, describe the pattern you expect to observe on the screen. Explain your reasoning."

Q1. Situation described above.

Q2. Vertical polarizer placed in front of one slit.

Q3. Vertical polarizer placed in front of each slit.

Q4. Vertical and horizontal polarizer placed in front of slits 1 and 2, respectively.

Q5. Vertical and horizontal polarizer placed in front of slits 1 and 2, respectively. Additionally, a polarizer which makes an angle of $+45°$ with the horizontal is placed in between the slits and the screen.

These questions are isomorphic to questions students considered in the context of the MZI: Q1—no polarizers placed in either path of the MZI, Q2—vertical polarizer placed in one path of the MZI, etc. The MZI pre- and post-tests were comprised of the MZI polarizer questions and many other questions in other situations (e.g., effect of removing BS2 on interference at the detectors D1 or D2 and the probabilities of photons of a given polarization arriving at D1 and D2 in different situations) which are very different from the polarizer questions.

## C. Research questions and study design

We investigated three research questions. The first two are related to the MZI QuILT and are referred to as RQ1.a and RQ1.b and the other is related to the DSE QuILT and is referred to as RQ2. We describe the research questions and the approach used to investigate them below and summarize them in Figs. 1 and 2 (RQ1.a and RQ1.b are summarized in Fig. 1 and RQ2 in Fig. 2).

We hypothesized that at least some students who learn how to reason about the MZI polarizer questions in terms of WPI from the MZI QuILT may be able to use this reasoning appropriately when answering the DSE polarizer questions which were isomorphic to the MZI polarizer questions. If students are indeed able to use WPI reasoning when answering the DSE polarizer questions, this would be an indication that the MZI QuILT may be effective at promoting WPI reasoning.

Thus, we first wanted to determine the percentage of students who use WPI reasoning on the MZI and DSE polarizer questions before having the opportunity to learn from the MZI QuILT and compare to after they work on the MZI QuILT (this is the focus of RQ1.a).

(We note that in all that follows, "MZI pretest" refers to MZI questions given before the MZI QuILT, "DSE pretest" refers to DSE questions given before the DSE QuILT, and "DSE post-test" refers to DSE questions given after the DSE QuILT. We have described student performance and use of appropriate reasoning on the MZI post-test elsewhere [37], and thus, this is not discussed here.)

*RQ1.a. What percentage of students use WPI reasoning before working on the MZI QuILT and how does that compare to after working on the MZI QuILT?*

To investigate this question, we gave the MZI pretest, which includes the MZI polarizer questions, to 46 undergraduate and 45 graduate students before these students worked on the MZI QuILT. We refer to these students as the MZI → DSE cohort because they worked on the MZI QuILT (pretest, tutorial, post-test) before working on the DSE QuILT. Thus, these students did not have an opportunity to learn from the MZI QuILT before answering the MZI polarizer questions. After they answered the MZI polarizer questions, these students worked on the MZI QuILT and then answered the DSE polarizer questions. We compared the percentage of students who used WPI reasoning after working on the MZI QuILT to before working on the MZI QuILT.

A second cohort of 14 graduate students (referred to as the DSE → MZI cohort) worked on the DSE QuILT first (pretest, tutorial, post-test) before they worked on the MZI QuILT. Thus, these students did not have the opportunity to learn about WPI from the MZI QuILT before answering the DSE polarizer questions. We compared the percentage of graduate students from the DSE → MZI cohort who used WPI reasoning to answer the DSE polarizer questions with the percentage of graduate students from the MZI → DSE cohort who used WPI on these questions.

Furthermore, if the MZI QuILT is effective in promoting the use of WPI reasoning, students who work on the MZI QuILT before answering the DSE polarizer questions should perform better on these questions compared to students who do not work on the MZI QuILT before answering them. However, for the DSE lamp questions, which do not have isomorphic pairs discussed in the MZI QuILT, we should observe no statistically significant differences in the performance between students who work on the MZI QuILT before answering these questions compared to students who do not work on the MZI QuILT before answering them. This is the focus of RQ1.b below.

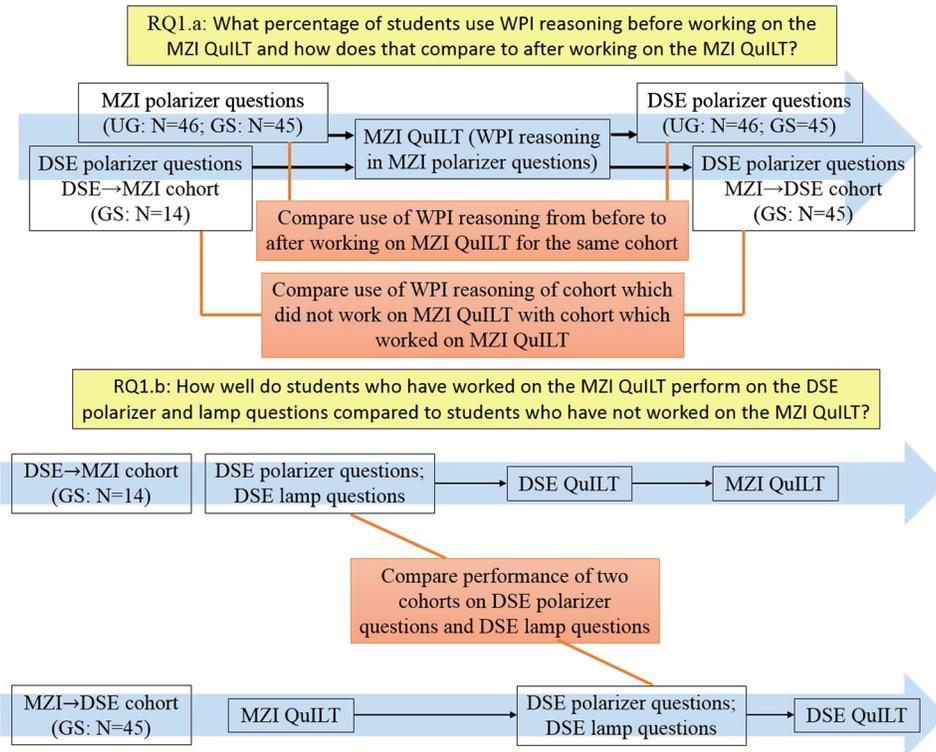

FIG. 1.   Schematic description of the design used to investigate RQ1.a and RQ1.b. Abbreviations in the figure are undergraduate students (UG), graduate students (GS), and $N$ refers to the number of students. The timeline of chronological order is from left to right as depicted by the blue arrows. All of the questions were given as pretests, i.e., the DSE polarizer questions in both RQ1.a and RQ1.b were always given before students worked on the DSE QuILT, and the MZI polarizer questions in RQ1.a were given before students worked on the MZI QuILT.

*RQ1.b. How well do students who have worked on the MZI QuILT perform on the DSE polarizer and lamp questions compared to students who have not worked on the MZI QuILT?*

To investigate this question, for the two graduate student cohorts who either worked or did not work on the MZI QuILT before answering the DSE pretest questions, we used a one way repeated measures ANOVA [38] to investigate the interactions between the conditions (working on the MZI QuILT vs not working on the MZI QuILT) and two performance measures (DSE lamp questions and DSE polarizer questions). We hypothesized that if the MZI QuILT is effective in promoting WPI reasoning, we would find that the only significant interaction is between the condition of working on MZI QuILT and performance on the DSE polarizer questions.

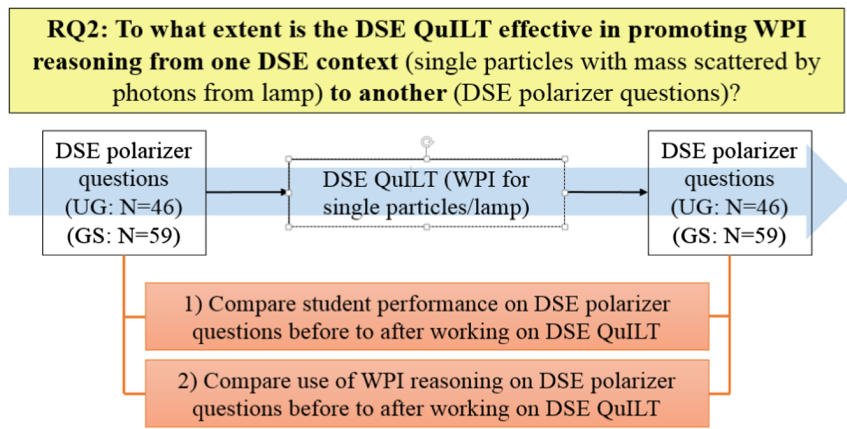

FIG. 2.   Schematic description of the design used to investigate RQ2. UG, GS, and $N$ have the same meaning as in Fig. 1. The timeline of chronological order is from left to right as depicted by the blue arrow.

The study design for RQ1.a and RQ1.b including chronology and number of participants is summarized in Fig. 1.

*RQ2. To what extent is the DSE QuILT effective in promoting WPI reasoning from one context of the DSE (single particles and a monochromatic lamp placed between the slits and the screen) to a different context of the DSE (single photons and polarizers placed in front of one or both slits) without an instructional intervention designed to help them make the connection between these different contexts?*

Students in all cohorts worked on the DSE pretest, and then worked on the DSE QuILT after which they worked on the DSE post-test. Some of the students worked on the MZI QuILT before working on the DSE (pretest, tutorial, post-test), while other students worked on the MZI QuILT after working on the DSE. We therefore investigated the extent to which the DSE QuILT, which discussed WPI reasoning in the context of the DSE with single particles with mass and a monochromatic lamp placed between the slits and the screen, promoted WPI reasoning to a different context (DSE polarizer questions) by comparing their performance and use of WPI reasoning before working on the DSE QuILT to after working on it. The study design for RQ2 including chronology and number of participants is summarized in Fig. 2.

The summary of the rubric used to grade students' performance on the DSE polarizer questions is shown in Table I. This rubric is designed to evaluate students' conceptual understanding of the effect of placing polarizers of various orientations in front of one or both slits in the DSE by considering responses for multiple questions together. For example, the third conceptual point (recognize that

"which-path" information can be lost) is based on students' answers to the last two questions (orthogonal polarizers and quantum eraser). Similarly, the second conceptual point is based on students' answers to all the questions: students should recognize that the situation in which two polarizers are orthogonal and there is no quantum eraser is the only case in which no interference is observed on the screen. If students claimed that the interference pattern vanishes in more than one situation, it was considered that they did not understand this conceptual point.

Another rubric was designed for grading the DSE lamp questions with the same goal: assess student understanding of concepts across questions. However, discussing this rubric in detail here would require discussing those questions along with the correct answers, which we have done elsewhere [34].

The two rubrics were used to score students' performance on the DSE polarizer questions and DSE lamp questions. A subset of the responses for all questions (20%–30%) was graded separately by two investigators. After comparing the grading of some students, the raters discussed any disagreements in grading and resolved them so that the interrater agreement after the discussions was better than 90%.

Before discussing the results, we note that we report the following in the results section:

- The percentage of students who used WPI reasoning among those who wrote down any reasoning (Tables II and VI).
- The percentage of students who answered each of the DSE polarizer questions correctly (Table III). Students who did not answer a particular question were excluded from the data analysis for that question. However, students were given more than enough time to complete the DSE pretest, and nearly all students handed in their pretests voluntarily. In Appendix C, we provide statistics for how many students did not provide a response on each of these questions.
- For student performance on a group of questions, in particular, the DSE polarizer questions and the DSE lamp questions (data shown in Table IV and Table V), students' performance was graded using rubrics (the summary of the rubric used for the DSE polarizer questions is provided in Table I.

TABLE I. Summary of the rubric used to grade students' performance on the DSE polarizer questions.

| | |
|---|---|
| Recognize that photons exhibit interference | +1, 0 |
| Recognize that only[a] when two polarizers are orthogonal and there is no "quantum eraser"—the interference pattern vanishes | +1, 0 |
| Recognize that "which-path" information can be lost | +1, 0 |
| Correctly interpret the effect of one polarizer on the interference pattern | +1, 0 |
| Correctly interpret the effect of two polarizers on the interference pattern (both questions) | +2, 1, 0 |

[a]if a student said that the interference pattern vanishes in more than 1 situation → 0 points.

TABLE II. Percentages of undergraduate students (US) and graduate students (GS) who used WPI reasoning out of those who provided reasoning on DSE polarizer questions 2-5 (Q2-Q5) in the DSE pretest. All these students worked on the MZI QuILT prior to taking the DSE pretest. We note that on average, 67% of the undergraduate students and 28% of the graduate students provided reasoning for their answers. There data are based on 46 undergraduate and 45 graduate students.

| | Q2 | Q3 | Q4 | Q5 |
|---|---|---|---|---|
| US | 37 | 37 | 57 | 62 |
| GS | 33 | 20 | 60 | 44 |

TABLE III. Percentage of graduate students from the DSE → MZI cohort and the MZI → DSE cohort who answered DSE polarizer questions Q1 through Q5 correctly on the DSE pretest. In the table, $N$ refers to the number of graduate students in each cohort.

|  | $N$ | Q1 | Q2 | Q3 | Q4 | Q5 |
|---|---|---|---|---|---|---|
| DSE → MZI cohort | 14 | 77 | 8 | 58 | 80 | 50 |
| MZI → DSE cohort | 45 | 88 | 46 | 78 | 81 | 71 |

The statistical tests we used to analyze the data are the following:

- For the data shown in Table V, we used a simple $t$ test for both populations because we were comparing the performance of each group (undergraduate and graduate students) from before to after working on the DSE QuILT. We also report effect sizes (Cohen's $d$ [38]) for the improvement in performance on the DSE polarizer questions from before to after working on the DSE QuILT. The guidelines for interpreting Cohen's $d$ are that a value of 0.2 corresponds to a small effect, a value of 0.5 to a medium effect and a value of 0.8 to a large effect [38].

- For the data in Table 6 $p$ values were obtained by conducting MacNemar's test, and the effect sizes we report are Cramer's V (equivalent to phi[1] when 2 groups are being compared) [38]. The general guidelines for Cramer's V (when two groups are being compared) is that a value of 0.1 corresponds to a small effect, 0.3 to a medium effect and 0.5 to a large effect [38].

- For the data in Table IV, we used a one way repeated measures ANOVA [38] to investigate the interactions between condition (working on the MZI QuILT vs not working on the MZI QuILT) and two performance measures (lamp questions and polarizer questions).

Also, in all cases for which undergraduate performance or percentage correct is shown, the data are based on all 46 undergraduate students. For graduate students, there are either 45 or 14 depending on the cohort (45 in the MZI → DSE cohort, 14 in the DSE → MZI cohort).

---

[1]Phi, or the phi coefficient, defines the strength of the relationship described in a 2 × 2 contingency table (in other words, phi is a measure of the effect of the difference between two groups for nominal data, i.e., students who used WPI reasoning, students who didn't use WPI reasoning before and after working on the DSE QuILT; in our case, phi is a measure of the effect size when comparing the percentage of undergraduate/graduate students who used WPI reasoning before working on the DSE QuILT to after working on the DSE QuILT). Cramer's V is an extension of the phi coefficient for contingency tables with more than 2 rows and 2 columns (i.e., similar to how ANOVA is an extension to a $t$ test when more than 2 groups are being compared). For more information see Ref. [38].

TABLE IV. Performance of two graduate student cohorts (depending on the order in which they worked on the MZI and DSE tutorials) on the DSE transfer questions and on the DSE lamp questions in the DSE pretest. The $p$ values were obtained by carrying out a one-way repeated measures ANOVA to identify the significance of the interactions between condition (MZI → DSE cohort or DSE → MZI cohort) and two performance measures (DSE lamp questions or DSE polarizer questions). We also report Cohen's $d$ effect sizes to compare the performance of the two different cohorts. Std. dev. refers to standard deviation.

|  | MZI → DSE cohort ($N=45$) | | DSE → MZI cohort ($N=14$) | | | |
|---|---|---|---|---|---|---|
|  | Average | Std. dev. | Average | Std. dev. | $p$ | Cohen's $d$ |
| DSE polarizer questions | 65% | 13% | 38% | 26% | 0.011 | 0.831 |
| DSE lamp questions | 42% | 34% | 42% | 27% | 0.955 | 0.016 |

TABLE V. Performance of undergraduate (UG) and graduate students (GS) on the DSE polarizer questions before and after working on the DSE QuILT and $p$ values comparing the performance before to after working on the DSE QuILT for each group (obtained via a simple $t$ test) as well as effect sizes (Cohen's $d$). $N$ refers to the number of undergraduate or graduate students. Std. dev. refers to standard deviation.

|  |  | DSE polarizer questions | |
|---|---|---|---|
|  | $N$ | Average | Std. dev. |
| UG-before QuILT | 46 | 57% | 35% |
| UG-after QuILT | 46 | 88% | 20% |
| UG $p$ value | | <0.001 | |
| UG effect size | | 1.09 | |
| GS-before QuILT | 59 | 60% | 35% |
| GS-after QuILT | 59 | 75% | 28% |
| GS $p$ value | | 0.018 | |
| GS effect size | | 0.47 | |

## IV. RESULTS

*RQ1.a. What percentage of students use WPI reasoning before working on the MZI QuILT and how does that compare to after working on the MZI QuILT?*

For the MZI polarizer questions, we found that without any prior instruction related to WPI, only one graduate student (out of 45) used WPI reasoning and only to answer one question. None of the 46 undergraduate students used this reasoning. For the DSE polarizer questions, we also found that prior to any instruction on WPI (from the MZI QuILT), only one graduate student (out of 14) used WPI reasoning to answer only one question. Thus, prior to any instruction on WPI, almost none of the graduate or undergraduate students use WPI reasoning to answer questions about interference on the MZI or DSE polarizer questions.

Table II shows, for DSE polarizer questions 2–5, the percentages of both undergraduate and graduate students (from the MZI → DSE cohort) who used WPI reasoning out of the students who provided any reasoning for their answers on the DSE pretest. All these students worked on the MZI QuILT before answering the DSE polarizer questions, but had not worked on the DSE QuILT. We note that despite the fact that all questions explicitly asked for reasoning, some students did not provide any reasoning. On average, 67% of the undergraduate students and 28% of the graduate students provided reasoning for their answers on the DSE polarizer questions in the DSE pretest. However, it is possible that some of the students who did not explicitly write down their reasoning may have answered the questions by reasoning about WPI, and if that was the case, that would imply these students learned how to use WPI reasoning from the MZI QuILT, just that they did not explicitly provide their reasoning. Table II shows that students who worked on the MZI QuILT before answering the DSE polarizer questions on the DSE pretest often used WPI reasoning to answer these questions, especially on the last two questions. In addition, out of all instances in which a graduate or undergraduate student used WPI reasoning and described their reasoning to answer a question, he/she used it correctly 79% of the time, thus indicating appropriate usage of WPI reasoning learned in the MZI context to answer questions in the DSE context. In contrast, as mentioned earlier, students almost never used such reasoning when answering the MZI or DSE polarizer questions when they did not have the opportunity to learn about the concept of WPI from the MZI QuILT. This suggests that students' use of the WPI reasoning on the DSE polarizer questions may be due to recognizing how to use this reasoning which they learned in the MZI QuILT to answer questions in a different context (DSE).

*RQ1.b. How well do students who have worked on the MZI QuILT perform on the DSE polarizer and lamp questions compared to students who have not worked on the MZI QuILT?*

Table III shows the percentage of graduate students from the DSE → MZI and MZI → DSE cohorts who answered DSE polarizer questions Q1 through Q5 correctly on the DSE pretest. Although the numbers are too small to perform meaningful statistics on each individual question, Table III suggests that students who had the opportunity to learn from the MZI QuILT were more likely to answer these questions correctly (meaningful statistics can, however, be performed on the aggregate data, i.e., overall performance on the DSE polarizer questions; see Table IV and the discussion related to the data shown in Table IV).

Table IV shows the average performance of the two graduate student cohorts on the DSE polarizer questions (as graded using the rubric shown in Table I), as well as their performance on the DSE lamp questions, which were quite different from the DSE polarizer questions. Students in the MZI → DSE cohort worked on the MZI QuILT before answering these questions whereas students in the DSE → MZI cohort did not. A repeated measures ANOVA carried out on these data showed a statistically significant interaction between working on the MZI QuILT and performance on the polarizer questions ($F(1, 51) = 7.034$, $p = 0.011$). The other interaction (working on MZI QuILT and lamp questions) was not significant, thus suggesting that the MZI QuILT helped students on the DSE polarizer questions only. Furthermore, the effect size [38] for comparing the performance on the DSE polarizer questions of graduate students who worked on the MZI QuILT with the performance of those who did not was 0.83, thus suggesting a large effect of working on the MZI QuILT on these questions.

*RQ2. To what extent is the DSE QuILT effective in promoting WPI reasoning from one context of the DSE (single particles and a monochromatic lamp placed between the slits and the screen) to a different context of the DSE (single photons and polarizers placed in front of one or both slits) without an instructional intervention designed to help them make the connection between these different contexts?*

Table V shows the overall performance of undergraduate and graduate students (averages and standard deviations) on the DSE polarizer questions as graded by the rubric shown in Table I both before working on the DSE QuILT and after (all graduate students from both cohorts are included in the graduate student data). Table V also lists $p$ values obtained via a simple $t$ test and effect sizes (Cohen's $d$ [38]) comparing students' performance from before to after working on the DSE QuILT. The $p$ values show that both undergraduate and graduate students improved significantly. The effect is large for undergraduate students, but only medium for graduate students. The improvement may seem surprising because the DSE QuILT did not address any of the situations in the DSE polarizer questions at all, and did not even mention interference of *photons* in the DSE. We discuss some possible reasons for this improvement in detail in Sec. V.

Table VI shows, for DSE questions 2–5, the percentages of both undergraduate and graduate students (similarly to Table V, all the graduate students are included in these data) who provided reasoning related to WPI among those who provided any reasoning for their answers both before and after working on the DSE QuILT. As mentioned earlier, students did not always provide reasoning for their answers even though the questions explicitly asked for reasoning. The percentages of both undergraduate and graduate students who provided reasoning on the DSE polarizer questions before working on the DSE QuILT were mentioned earlier in this section (under *RQ1.a*); after working on the DSE QuILT, on average 96% of the undergraduate students provided reasoning for their answers while on average, 49% of the graduate students provided reasoning for their answers. The $p$ values listed in Table VI show that undergraduate students were statistically significantly more likely to provide reasoning related to WPI after working on the DSE QuILT on three out of the four questions, and for the graduate students it was two out of the four questions. The effect sizes (Cramer's V) shown in Table VI suggest that for most questions the magnitude of the effect is medium. Given that students who used WPI reasoning used it correctly 79% of the time, it appears that increased usage of WPI reasoning may play an important role in the improvement observed in Table V for both graduate and undergraduate students.

TABLE VI. Percentage of undergraduate (UG) and graduate (GS) students who used WPI reasoning among those who provided reasoning on DSE polarizer questions 2–5 and $p$ values and effect sizes for comparing these percentages from before to after working on the DSE QuILT via MacNemar's tests.

|  | $N$ | Q2 | Q3 | Q4 | Q5 |
|---|---|---|---|---|---|
| UG-before QuILT | 46 | 37 | 37 | 57 | 62 |
| UG-after QuILT | 46 | 88 | 52 | 87 | 88 |
| $p$ value |  | <0.001 | 0.057 | 0.021 | 0.021 |
| Effect size |  | 0.62 | 0.22 | 0.36 | 0.30 |
| GS-before QuILT | 59 | 27 | 14 | 47 | 31 |
| GS-after QuILT | 59 | 48 | 52 | 77 | 70 |
| $p$ value |  | 0.070 | 0.016 | 0.073 | 0.013 |
| Effect size |  | 0.23 | 0.43 | 0.32 | 0.42 |

## V. DISCUSSION

As evidenced in Tables V and VI, both the graduate and undergraduate students exhibited improved performance on the DSE polarizer questions after working on the DSE QuILT, and they were also more likely to make use of WPI reasoning to motivate their answers (and most students who used WPI reasoning did so correctly). This improved performance on the DSE polarizer questions may seem surprising. However, the DSE QuILT did guide students through the concept of WPI and how it can be used to determine whether interference is observed in the DSE with single particles when a monochromatic lamp which emits photons that scatter with the particles (with mass) is placed between the slits and the screen. In some of these situations, scattering between the particles emitted by the source and the photons emitted by the lamp can provide WPI for the particles and destroy the interference pattern. It is possible that students who engaged with the DSE QuILT deeply can recognize on their own how this type of WPI reasoning can be applied to answer the DSE polarizer questions.

To test this hypothesis we conducted think-aloud interviews with students who had completed the study of Modern Physics 1, which (typically) discusses the basic set up of the DSE. In an interview, students answered the DSE pretest questions, worked on the DSE QuILT, and then answered the DSE post-test questions while thinking aloud. We note that these students had not worked on the MZI QuILT so there was no possibility of transfer of the WPI concept and its relation to interference from the MZI context to the DSE context. Students were not disturbed during the interviews except when they became quiet for a long time, in which case the interviewer prompted the student to keep talking. After working on each part (e.g., pretest), students were asked for clarification on points they had not made clear earlier while talking aloud. We also should stress that this qualitative investigation is not the primary focus of this investigation and thus the data collected was not analyzed in great detail, e.g., by transcribing and coding it, but rather the interviewer paid close attention to how students were guided by the DSE QuILT and took careful notes during the interviews.

The interviews suggested that the DSE QuILT helped students reason using WPI to determine the pattern observed on the screen for a given DSE setup. In many cases, they were able to transfer this reasoning correctly to the DSE polarizer questions. For example Andrew, one interviewed student, when answering DSE polarizer question 3 (a vertical polarizer placed in front of each slit) before working on the DSE QuILT noted that a full interference pattern will form, however, he was not sure why. When the interviewer probed further (after the student had answered all pretest questions) it appeared that the student was primarily guessing on this question and he did not have a very good reason for his answer. On the other hand, after working on the DSE QuILT, when answering the same question he said: "There will be interference. If the photon is vertical (vertically polarized), there is no which path knowledge, so there is interference. If (the photon is) horizontal, it doesn't go through."

Thus, Andrew reasoned correctly using the concept of WPI, which he had learned in the DSE QuILT in completely different situations involving placing a monochromatic lamp between the slits and the screen for a DSE with single particles with mass instead of using single photons and placing polarizers of various orientations in front of one or both slits (polarizer questions). After working on the DSE QuILT, Andrew used WPI reasoning to answer the other DSE polarizer questions, and for the most part, used this reasoning correctly. For example, on DSE polarizer question 4 (two orthogonal polarizers) he recognized that WPI is known for all photons and therefore no interference is observed on the screen.

John, another interviewed student, while working on DSE polarizer question 4 before completing the DSE QuILT, understood that the vertically polarized photons will go through one slit and the horizontally polarized photons will go through the other. However, he thought that both will create an interference pattern. He stated: "So there are two cases to consider: one where there's (…) a horizontal photon coming in and the other is when there's a vertical photon coming in. So if it's a horizontal photon coming in, it only goes through the right one [slit with horizontal polarizer] and you get an [interference] pattern, and if the vertical one [photon] comes in, it only goes through the left one and you get an [interference] pattern. I don't know if those patterns are going to overlap (…) If they overlap you'd just get a normal [interference] pattern, but if they don't overlap, you'd get a continuum [random background]"

Discussions suggest that initially John thought that both the horizontally and the vertically polarized photons will create an interference pattern, and depending on where the two patterns form, they can either overlap perfectly, or are offset by a half of a wavelength so that the highs of one pattern overlap over the lows of the other pattern to produce an overall homogeneous distribution.

On the other hand, after working on the DSE QuILT, John correctly reasoned that a horizontally and a vertically polarized photon each goes through only one slit, and therefore no interference pattern is observed because WPI is known. In all the questions with polarizers, he reasoned by thinking about WPI, which is a concept he learned in the DSE QuILT in a different context. Interestingly, when reading the first DSE polarizer question in the post-test he stated "Hmm… So I don't think this was in the tutorial, but I assume something in the tutorial should help me answer these (questions)". It appeared that he was able to use what he learned about how gaining WPI affects the pattern observed on the screen to reason about the DSE polarizer questions. It is possible that similar reasoning applies to other students like John who improved on the DSE polarizer questions after working on the DSE QuILT, which did not discuss the setups in the DSE polarizer questions.

It is important to keep in mind that these students only worked on the DSE QuILT and were not exposed to the MZI QuILT at all. It appears that they were able to make connections between what they learned in the DSE QuILT, in particular how to reason in terms of WPI to determine whether an interference pattern is formed, to answer the DSE polarizer questions. It is possible that if they had also worked on the MZI QuILT earlier, they would have been able to make connections between the type of WPI reasoning used in the MZI context and similar reasoning used in the DSE context. In that case, working on both tutorials is likely to consolidate their knowledge of WPI further and can lead to even better performance on the post-test, similar to the undergraduate students for whom both tutorials were a part of their course, as shown in the results section. We note that the interviews provide a good starting point for understanding possible reasons for the QuILTs promoting WPI reasoning, and future investigations will probe these issues further.

Finally, we should note that even though the DSE QuILT did not discuss interference of single photons, since students answered the DSE polarizer questions as part of the DSE pretest, it is possible that this may have primed them to think about the DSE polarizer questions while working on the DSE QuILT, and this may have partly helped them make connections between what they were learning in the DSE QuILT (using WPI reasoning to explain interference for particles with mass) and the DSE polarizer questions. While this may have aided them, we also note that giving a pretest before instruction in a particular topics and then giving the identical post-test is a common practice in introductory physics (for example, giving the Force Concept Inventory [39] before and after instruction). However, in the context of introductory physics, it has been found that giving the FCI as a pretest does not bias post-test results [40].

## VI. SUMMARY

In this study, we find evidence that a QuILT on the MZI was effective in promoting WPI reasoning from the MZI context to help upper-level undergraduate and graduate students answer isomorphic questions in the context of the DSE. The MZI QuILT introduced students to the concept of WPI and guided them to use this concept to reason about whether or not interference is observed at the detectors in a particular MZI setup. Among students who did not work on the MZI QuILT, almost none of them made use of WPI reasoning when answering either the MZI or DSE polarizer questions. In contrast, after working on the MZI QuILT, the percentages of students who used WPI reasoning on the DSE polarizer questions (among those who provided reasoning) ranges from 20% to 60% for the graduate students and 37% to 62% for the undergraduate students. Additionally, the graduate students who worked on the MZI QuILT before answering the DSE polarizer questions on the pretest performed significantly better on these questions than the graduate students who did not. These two cohorts of graduate students showed identical performance on the other DSE questions, which did not have analogous situations discussed in the MZI QuILT, suggesting that the improved performance on the DSE polarizer questions is likely due to the MZI QuILT helping students discern the underlying principles required to answer questions about interference by using WPI reasoning. We note however that the number of graduate students the DSE → MZI cohort was small (14), and thus it is possible that the encouraging results can be at least in part accounted for by the small number of students. We also note that due to lack of participation from faculty members teaching the upper level undergraduate quantum mechanics course, we were unable to investigate the DSE → MZI condition for undergraduate students. However, our research with the DSE QuILT [34] and the MZI QuILT [37] indicated that the undergraduate students learned more from these QuILTs compared to graduate students. In both the DSE and MZI, on the pretest, undergraduate students' performance was lower on average than graduate students' performance, but on the post-test, undergraduate students outperformed graduate students, thus indicating higher normalized gains for the undergraduate students. In this study too, we found that undergraduate students' performance on the DSE polarizer questions improved more than graduate students' performance—data shown in Table V. We hypothesize that if we had investigated the DSE → MZI condition for undergraduate students, we may have found similar, if not more encouraging results.

In addition, we found that after working on the DSE QuILT, both undergraduate and graduate students improved significantly on the DSE polarizer questions despite the fact that the DSE QuILT did not mention anything about polarizers thus indicating that the DSE QuILT also promoted WPI reasoning from one context to another. Interviews with students who worked *only* on the DSE QuILT suggest that this improved performance may partly be due to students correctly recognizing the utility of WPI reasoning when considering situations described in the DSE polarizer questions. It is likely that students who work on the MZI QuILT before working on the DSE QuILT and engage with both tutorials well, e.g., the undergraduates who worked on both QuILTs as part of their quantum mechanics course, consolidate their knowledge of WPI further by making connections between the DSE and MZI contexts.

Finally, we note that the results of this investigation can be interpreted from the lens of transfer of learning [20,41–51]. In other words, the QuILTs were effective in promoting positive transfer of learning [48] of WPI reasoning, and advanced students recognized how to apply the WPI concept learned in one context to a different context on their own without an instructional intervention designed to help them make the connection between the different contexts. The results reported here show evidence of significant positive transfer, something that has rarely been found in prior research. However, most prior studies in knowledge-rich domains such as physics have focused on transfer among introductory students who have significantly less prior relevant knowledge and skills compared to advanced physics students that may be crucial for positive transfer. In other words, it is possible that part of the reason for transfer found here is that this investigation has been carried out with advanced undergraduate students and graduate students who have a greater level of relevant prior knowledge and problem solving, reasoning, and metacognitive skills which may play a role in facilitating transfer [52]. The effectiveness of the research-based tutorials in positively transferring learning from one context to another in advanced courses is promising and is useful for researchers investigating positive transfer of learning at all levels. Future studies focusing on the reasons for positive transfer in advanced courses similar to the one discussed here can further help shed light on the nature of expertise in advanced courses and how the prior knowledge and skills of the learner and the features of the research-based curricula may interact to enable positive transfer.

## ACKNOWLEDGMENTS

We thank the National Science Foundation for Grant No. PHY-1505460, Albert Huber for creating the simulation used in the MZI QuILT, and Klaus Muthsam for creating the simulation used in the DSE QuILT. We also thank F. Reif, R. P. Devaty, and E. Marshman for helpful discussions.

## APPENDIX A: ISOMORPHISM BETWEEN MZI AND DSE QUESTIONS

Since WPI reasoning can be used to reason about interference in the MZI and DSE, we begin by defining WPI: in general, when a detector can project both components of the path state, then WPI is unknown. On the other hand, when a detector can project only one component of the path state, then we have complete which path information, or WPI is known.

We first consider the most basic MZI setup shown in Fig. 3. BS1 and BS2 are beam splitters; BS1 is oriented such that it puts the single photon emitted from the source into an equal superposition of the $U$ and $L$ path states shown (which we represent as $|U\rangle$ and $|L\rangle$, respectively). Mirrors are for proper alignment. BS2 ensures that the components of the single photon state from both the $U$ and $L$ paths can be projected into each (photo) detector D1 and D2 after BS2 so that constructive or destructive interference (or anything in between) can be observed (depending on the path length difference between the $U$ and $L$ paths). If an *additional* detector is placed anywhere in the lower path $L$ between BS1 and BS2, after encountering the detector, the superposition of the $U$ and $L$ path states of a photon collapses and if the photon does not get absorbed by the detector, the state of the photon inside the MZI is the upper path state $|U\rangle$. Conversely, if an additional detector is placed in the upper path $U$, after encountering the detector, if the photon is not absorbed by that detector, the state of the photon inside the MZI collapses to the lower path state $|L\rangle$. In these situations (additional detector in the $U$ or $L$ path of the MZI), if a photon arrives at the detector D1 or D2 after BS2, we have WPI because either detector can only project the component of the photon state along the U or L path and no interference is observed at D1 or D2. If instead, no detector is placed in either of the $U$ or $L$ path of the MZI (as in Fig. 3), the state of a photon inside the MZI remains an equal superposition of the $U$ and $L$ path states, WPI is unknown (because the detectors can project both the $|U\rangle$ and $|L\rangle$ components of the photon state) and therefore interference is observed at D1 and D2.

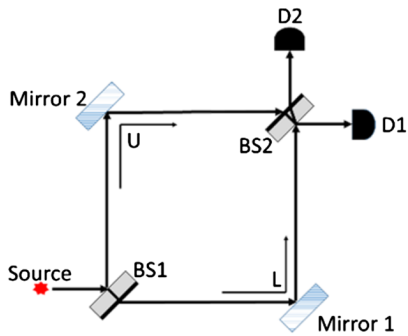

FIG. 3. Basic MZI setup.

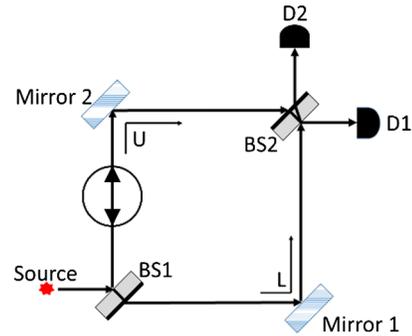

FIG. 5. MZI setup with a vertical polarizer placed in the upper path.

Now consider the DSE setup shown in Fig. 4. If slit 2 is blocked, the state of a photon inside the DSE (after passing through the slits) collapses to, e.g., $|\Psi_1\rangle$ and if slit 1 is blocked, the state of a photon collapses to, e.g., $|\Psi_2\rangle$. If this photon arrives at the screen (the screen is the detection device in the DSE, equivalent to detectors D1 and D2 in the MZI), we have WPI because the screen can only project one component of the photon's path state (either $|\Psi_1\rangle$ or $|\Psi_2\rangle$) and therefore, no interference is observed. If neither slit is blocked, the state remains an equal superposition of $|\Psi_1\rangle$ and $|\Psi_2\rangle$. In other words, $|U\rangle$ and $|L\rangle$ in the MZI are analogous to $|\Psi_1\rangle$ and $|\Psi_2\rangle$ in the DSE. In the situations in which there is no detector in either path of the MZI and neither slit is blocked for the DSE, we do not have WPI and each photon interferes with itself.

Now consider the situation shown in Fig. 5 in which we place a vertical polarizer in the upper path of the MZI and the source emits $+45°$ polarized single photons. This situation is analogous to the situation shown in Fig. 6 in the DSE in which a vertical polarizer is placed after slit 1 (and the source emits $+45°$ polarized single photons). We now have to use a four dimensional Hilbert space, two dimensions for path/slit states, $|U\rangle$,$|L\rangle$/$|\Psi_1\rangle$, $|\Psi_2\rangle$, and two dimensions for polarization states, for which the convenient basis for the situations described in Figs. 5 and 6 is $\{|V\rangle, |H\rangle\}$ (vertical, horizontal polarization states, respectively). If a vertical polarizer is placed in the upper path of the MZI, the $|U\rangle$ state will be associated with a vertical polarization state ($|U\rangle|V\rangle$) and the $|L\rangle$ state is still associated with both vertical and horizontal polarization states ($|L\rangle|V\rangle + |L\rangle|H\rangle$). In both experiments we assume that the detectors are sensitive to polarization (they are covered with polarizers with a particular orientation, e.g., vertical or horizontal), which means that the collapse of the photon state after it is measured by the detectors D1 or D2 provides information about the polarization of the photon. Therefore, in the situation depicted in Fig. 5, we have WPI for horizontally polarized photons arriving at D1 and D2 because the horizontal polarization is associated with the lower path state only—each detector can only project the $|L\rangle$ component of the state of a horizontally polarized photon. We do not have WPI for the vertically polarized photons because the vertical polarization is associated both with the upper and the lower path states—each detector can project both the $|L\rangle$ and $|U\rangle$ components of the state of a vertically polarized photon. The fact that we have WPI for horizontally polarized photons and we do not have WPI for vertically polarized photons implies that the photons that arrive at the detectors in the $|V\rangle$ state interfere and those in the $|H\rangle$ state do not. In the DSE, the situation is analogous (Fig. 6): if a vertical polarizer is placed after slit 1, horizontally polarized photons arriving at the screen will not interfere, while vertically polarized photons arriving at the screen will show interference.

It is important to note that while questions about interference in the DSE and MZI contexts are isomorphic,

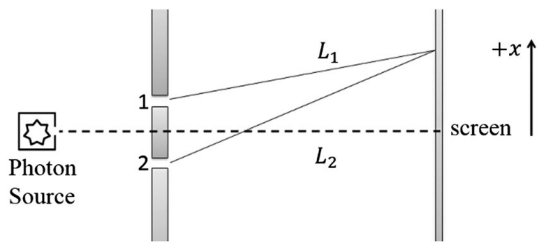

FIG. 4. Basic DSE setup with single photons.

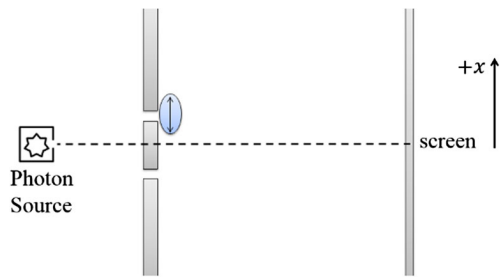

FIG. 6. DSE setup with a vertical polarizer placed after slit 1.

the "surface" features of these two experiments are rather different. In the MZI, the paths are restricted and the photons arrive at point detectors D1 and D2, while in the DSE the photons are delocalized in the space between the slits and the screen and can be detected anywhere on the extended screen. In addition, in the DSE, there is no explicit optical element corresponding to BS2 in the MZI which mixes the components of the photon state from the two paths. These differences suggest that the surface features of these problems are quite different, which can make it challenging for novices to recognize the isomorphism. In order to recognize the isomorphism between the MZI and DSE questions, students must be able to reason about the deep features of the contexts and recognize the utility of the concept of WPI and its relation to whether or not interference will take place in both contexts. Thus, even if students fully understand the underlying physics principles in the MZI context, they may have difficulty recognizing how the same physics principles apply to the DSE.

Also, it is worthwhile to keep in mind that while the upper-level undergraduates and graduate students have some knowledge of the DSE, almost none of them have been introduced to the concept of WPI and learned how to reason using WPI to answer questions similar to the ones discussed here. We found that among the graduate students who were not introduced to WPI via our tutorials, only one used WPI reasoning to answer only one question (out of five) on the DSE with polarizers placed in front of one slit. Similarly, in the MZI context before being introduced to WPI reasoning via tutorials, we also found that only one graduate student used this reasoning. The physics undergraduate students in this study were almost all nearly at the end of the undergraduate curriculum (more than 80% were seniors) and the physics graduate students were all in their first year. Therefore, for the purposes of this study, the two populations (undergraduates and graduate students) are not very different in terms of background knowledge on the DSE and MZI.

## APPENDIX B: DSE PRETEST

Note that the DSE post-test was identical with some minor differences (e.g., electrons in one problem being replaced by Na atoms). Questions 4–8 are what we refer to as the "DSE lamp questions" and questions 9(i) through 9(v) are the DSE polarizer questions Q1 through Q5 discussed at length in the article.

For all questions, ignore relativistic effects. For all questions that ask about a double slit setup, assume that the screen can detect the particles used and that the distance from the slits to the screen is much larger than the distance between the slits.

For any constant, e.g., the mass of an electron or muon or Planck's constant, use the following values:
- 1 eV = $1.6 \times 10^{-19}$ J

- keV = kilo electron volt = $10^3$ eV, meV = milli electron volt = $10^{-3}$ eV
- 1 mm = $10^{-3}$ m, 1 $\mu$m = $10^{-6}$ m, 1 nm = $10^{-9}$ m, 1 pm = $10^{-12}$ m
- Planck's constant = $h = 6.6 \times 10^{-34}$ Js
- magnitude of elementary charge (on an electron or proton) = $e = 1.6 \times 10^{-19}$ C
- speed of light = $c = 3.0 \times 10^8$ m/s
- mass of electron = $9.1 \times 10^{-31}$ kg
- mass of neutron = mass of proton = $1.7 \times 10^{-27}$ kg
- mass of muon = $1.9 \times 10^{-28}$ kg
- mass of helium atom = $6.7 \times 10^{-27}$ kg
- mass of sodium atom = $3.8 \times 10^{-26}$ kg

### 1. Pretest

(1) What is the de Broglie relation? In one or two sentences, explain its significance.

(2) You are conducting a double-slit experiment in which you send a large number of nonrelativistic electrons of the same kinetic energy one at a time towards a double-slit plate. The slit width is 50 pm, the slit separation is 1 nm and the distance between the slits and the screen is 3 m.

   (i) If the wavelength of the electrons is 9 pm, describe the pattern you expect to observe on the screen after a large number of electrons have passed through. Explain your reasoning.

   (ii) Suppose the experiment is modified by using protons instead of electrons while all following parameters are held fixed: kinetic energy, slit width, slit separation, distance between slits and screen. How does the pattern change, if at all?

   (iii) Explain your reasoning for your answer in 2 (ii).

(3) Consider particles of sand, which can be approximated as spheres of a radius of about 1/10 of a millimeter.

   (i) Do you expect that a double slit experiment with well-chosen parameters would show an interference pattern?

   (ii) Explain your reasoning for your answer in 3 (i).

      In questions 4–8, assume that particles are sent one at a time from the particle source. Figure 7 shows a double-slit experiment which was modified by adding a lamp (light bulb) between the double slit and the screen (slightly off to the side so it is not directly in front of the slits).

- Assume that when the lamp is turned on, if scattering occurs between a particle used in the double-slit experiment and a photon from the lamp, this scattering occurs at the slits only.
- Assume that ALL the particles scattered by photons still reach the screen.

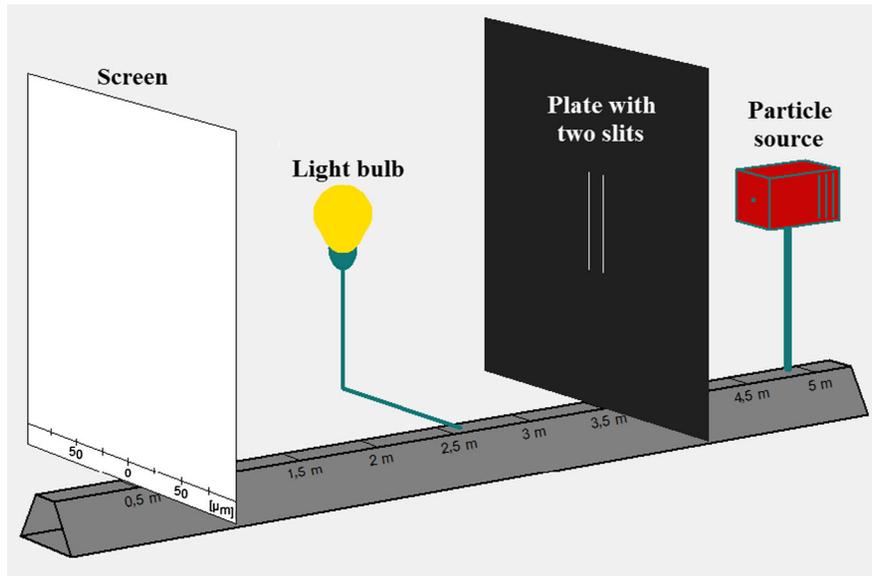

FIG. 7. Double slit setup with the addition of a lamp (image reproduced with permission from a simulation developed by Klaus Muthsam, muthsam@habmalnefrage.de).

(4) Suppose you perform a double slit experiment with electrons while the lamp is turned off and observe an interference pattern on the screen. You then repeat the experiment with the lamp turned on (assume that the intensity of the lamp is such that every particle used in the experiment scatters off a photon).

    (i) Describe a situation in which this addition of the lamp between the double slit and the screen destroys the interference pattern observed on the screen (in the situation you describe, assume that all particles reach the screen even if scattering occurs between the particles and the photons emitted by the lamp).

    (ii) Explain your reasoning for your answer in 4 (i).

Questions 5–8 refer to the following setup:

You perform a double-slit experiment using Na atoms and observe an interference pattern on the screen. You then change the experiment by adding a lamp as discussed earlier.

• If slit 2 is closed, the wave function of a Na atom that goes through slit 1 and arrives at a point $x$ on the screen is $\Psi_1(x)$. If instead, slit 1 is closed, the wave function of a Na atom that goes through slit 2 and arrives at a point $x$ on the screen is $\Psi_2(x)$.

• For this example, if slit 2 is closed, and a total number $N$ of particles arrives at the screen, the number density of the particles at a point $x$ on the screen is $N|\Psi_1(x)|^2$.

• For questions 5–8, both slits are open.

(5) For (i) and (ii) below, suppose that the wavelength of the photons is significantly <u>smaller</u> than the distance between the slits and the intensity of the lamp is such that <u>each Na atom scatters</u> off a photon. Also, assume that all the scattered atoms still reach the screen.

    (i) Write down an expression for the number density of Na atoms at a point $x$ on the screen in terms of $\Psi_1(x)$ and $\Psi_2(x)$ after a large number $N$ of Na atoms arrive at the screen.

    (ii) Describe the pattern you expect to observe on the screen after a large number $N$ of Na atoms have arrived at the screen. Explain your reasoning.

(6) For (i) and (ii) below, suppose that the wavelength of the photons is significantly <u>larger</u> than the distance between the slits and the intensity of the lamp is such that <u>each Na atom scatters</u> off a photon. Also, assume that all scattered atoms still reach the screen.

    (i) Write down an expression for the number density of Na atoms at a point $x$ on the screen in terms of $\Psi_1(x)$ and $\Psi_2(x)$ after a large number $N$ of Na atoms arrive at the screen.

    (ii) Describe the pattern you expect to observe on the screen after a large number $N$ of Na atoms have arrived at the screen. How, if at all, is this pattern different from the pattern in 5(ii)? Explain your reasoning.

(7) For (i) and (ii) below, suppose that the wavelength of the photons is significantly <u>smaller</u> than the distance between the slits and the intensity of the lamp is such that about <u>half of the Na atoms scatter</u> off a photon. Also, both slits are open and all the atoms reach the screen, including the ones that scatter.

    (i) Write down an expression for the number density of Na atoms at a point $x$ on the screen in terms of $\Psi_1(x)$ and $\Psi_2(x)$ after a large number $N$ of Na atoms arrive at the screen.

    (ii) Describe the pattern you expect to observe on the screen after a large number $N$ of Na atoms

have arrived at the screen. How, if at all, is this pattern different from the pattern in 5(ii)? Explain your reasoning.

(8) For (i) and (ii) below, suppose that the wavelength of the photons is significantly <u>larger</u> than the distance between the slits and the intensity of the lamp is such that about <u>half of the Na atoms scatter</u> off a photon. Also, both slits are open and all the atoms reach the screen, including the ones that scatter.

  (i) Write down an expression for the number density of Na atoms at a point $x$ on the screen in terms of $\Psi_1(x)$ and $\Psi_2(x)$ after a large number $N$ of Na atoms arrive at the screen.

  (ii) Describe the pattern you expect to observe on the screen after a large number $N$ of Na atoms have arrived at the screen. How, if at all, is this pattern different from the pattern in 6(ii)? Explain your reasoning.

(9) You perform a double slit-experiment in which photons that are polarized at $+45°$ are sent one at a time towards the double slit. The wavelength of the photons is comparable to the width of the slits and the separation between the slits is more than twice the slit width. <u>In all parts (i) through (vi) below, assume that the same large number $N$ of photons reach the screen</u> (in other words, you wait long enough in each case to clearly observe the pattern that forms on the screen).

  (i) Describe the pattern you expect to observe on the screen <u>after a large number $N$ of photons reach the screen</u>. Explain your reasoning.

  (ii) Suppose that a vertical polarizer is placed in front of only one of the slits. Describe the pattern you expect to observe on the screen <u>after a large number $N$ of photons reach the screen</u>. How does this pattern differ, if at all, from the pattern observed in 9(i)? Explain your reasoning.

  (iii) Suppose that a vertical polarizer is placed in front of each of the two slits. Describe the pattern you expect to observe on the screen <u>after a large number $N$ of photons reach the screen</u>. How does this pattern differ, if at all, from the pattern observed in 9(i)? Explain your reasoning.

  (iv) Suppose that a vertical polarizer is placed in front of one of the slits and a horizontal polarizer is placed in front of the other slit. Describe the pattern you expect to observe on the screen <u>after a large number $N$ of photons reach the screen</u>. How does this pattern differ, if at all, from the pattern observed in 9(i)? Explain your reasoning.

  (v) Suppose that a vertical polarizer is placed in front of one of the slits and a horizontal

polarizer is placed in front of the other slit. Furthermore, an additional polarizer which makes an angle of $+45°$ with the horizontal is placed right before the screen. Describe the pattern you expect to observe on the screen <u>after a large number $N$ of photons reach the screen</u>. How does this pattern differ, if at all, from the pattern observed in 9(i)? Explain.

# APPENDIX C: COMMON STUDENT DIFFICULTIES

Here, we discuss common student difficulties on the DSE polarizer questions both before and after students worked on the DSE QuILT. Since the data were qualitatively similar for the graduate students regardless of whether they had completed the MZI QuILT before taking the DSE pretest, the graduate students from all cohorts are combined. We also carried out think-aloud interviews with undergraduate and graduate students to further understand the common types of incorrect reasoning they used to answer these questions, which often provided further insight into their difficulties.

*Difficulties with interference of single photons—no polarizers*

Among the students who answered Q1, the vast majority of both undergraduate and graduate students answered it correctly (clear interference pattern shown) as shown in Table VII. A small percentage of students selected answers which indicated that no interference pattern is observed, but none provided reasoning for their answers. On the pretest, roughly one quarter of the undergraduate students and one sixth of the graduate students either did not respond or indicated that they did not know whether photons will exhibit interference in this case. These percentages drop to nearly zero in the post-test.

TABLE VII. Percentages of undergraduate (UG) and graduate students (GS) with different answers on Q1 (interference, no interference, other, and no response or "I don't know"). Bold italic indicates correct responses.

| | Interference | No interference | Other | No response/ "I don't know" |
|---|---|---|---|---|
| UG-before QuILT | ***63*** | 7 | 7 | 23 |
| UG-after QuILT | ***98*** | 2 | 0 | 0 |
| GS-before QuILT | ***71*** | 11 | 2 | 16 |
| GS-after QuILT | ***83*** | 5 | 8 | 3 |

TABLE VIII. Percentages of undergraduate (UG) and graduate students (GS) with different answers on Q2 (partial interference, no interference, full interference, other, and no response/"I don't know"). Bold italic indicates correct responses.

|                  | Partial interference | No interference | Full interference | Other | No response/"I don't know" |
| ---------------- | -------------------- | --------------- | ----------------- | ----- | -------------------------- |
| UG-before QuILT  | *38*                 | 17              | 5                 | 16    | 24                         |
| UG-after QuILT   | *70*                 | 16              | 2                 | 12    | 0                          |
| GS-before QuILT  | *31*                 | 19              | 12                | 21    | 16                         |
| GS-after QuILT   | *51*                 | 32              | 10                | 5     | 2                          |

### Difficulties with the effect of one polarizer on the interference pattern

Q2 involves a DSE in which a vertical polarizer is placed in front of only one of the slits. In this situation, WPI will be known for horizontally polarized photons and will not be known for vertically polarized photons (as explained in the section discussing the isomorphism between the DSE and MZI questions). Therefore, the pattern observed on the screen will consist of an interference pattern (neglecting single slit diffraction since the slit is sufficiently narrow) provided by the vertically polarized photons (which do interfere) on top of a uniform background provided by the horizontally polarized photons which do not interfere. This was the most challenging question for both student populations. As shown in Table VIII, for both populations, the most common incorrect answer choice is that no interference is observed in this situation. Students with this answer typically reasoned that WPI is known for all photons because the polarizer "tags" the photons that go through it by polarizing them (this reasoning did not always mention WPI explicitly). For example, one student stated: "no interference because you are essentially 'tagging' half the photons" and another stated "no interference since the polarizer tells us which slit the photon went through." This difficulty is also common in the MZI context when a vertical polarizer is placed in one of the paths: many students thought that no interference is observed at either detector because the polarizer provides WPI for the photons that take that path by 'tagging' them.

Interestingly, more graduate students use this type of reasoning after working on the DSE QuILT than before. This may be because before working on the DSE QuILT, some students (21%) provided responses that were difficult to categorize, and some (16%) did not provide a response, but after working on the DSE QuILT, the majority of these students provided responses that could be categorized, some of which used the incorrect reasoning that the vertical polarizer provides WPI for vertically polarized photons detected at the screen.

For students who attempted to explicitly reason in terms of WPI on the DSE polarizer questions, 67% of them (including both undergraduate and graduate students) reasoned correctly (note that this is the most challenging question for both undergraduate and graduate students). For example, one student wrote "The interference pattern will be fuzzier because we do have which-path data for any photons that are not vertically polarized" (common correct reasoning) and another wrote "I only see two lines on the screen because we have which-path information about one of the slits." The second student is using WPI reasoning incorrectly, but at the very least, he is recognizing that this reasoning may be useful in the DSE context.

### Difficulties with the effect of two polarizers on the interference pattern

Q3 and Q4 evaluate student understanding of the effect of two polarizers on the interference pattern. Students showed significant improvement after working on the DSE QuILT on these two questions as shown in Tables IX and X. Among the students who answered these questions before working on the DSE QuILT, the majority of them answered them correctly. Also, on these questions, the performance of undergraduate students after working on the DSE QuILT is close to 100%. It appears that the undergraduate students were able to use the concept of WPI learned from the MZI QuILT to answer the DSE polarizer questions, and also, after working on the DSE QuILT, they were able to consolidate their learning to develop a solid understanding of the effect of two polarizers on the interference pattern in the DSE. On the other hand, graduate students showed smaller improvement.

TABLE IX. Percentages of undergraduate (UG) and graduate students (GS) with different answers on Q3 (full interference, partial interference, no interference, other, and no response/"I don't know"). Bold italic indicates correct responses.

|                  | Full interference | Partial interference | No interference | Other | No response/"I don't know" |
| ---------------- | ----------------- | -------------------- | --------------- | ----- | -------------------------- |
| UG-before QuILT  | *52*              | 2                    | 14              | 6     | 26                         |
| UG-after QuILT   | *93*              | 2                    | 2               | 0     | 2                          |
| GS-before QuILT  | *60*              | 3                    | 14              | 5     | 17                         |
| GS-after QuILT   | *83*              | 7                    | 5               | 2     | 3                          |

TABLE X.   Percentages of undergraduate (UG) and graduate students (GS) with different answers on Q4 (partial interference, no interference, full interference, other, and no response/"I don't know"). Bold italic indicates correct responses.

|  | Full interference | Partial interference | No interference | Other | No response/"I don't know" |
|---|---|---|---|---|---|
| UG-before QuILT | 10 | 0 | *52* | 2 | 36 |
| UG-after QuILT | 2 | 0 | *93* | 0 | 5 |
| GS-before QuILT | 9 | 0 | *64* | 7 | 21 |
| GS-after QuILT | 8 | 7 | *81* | 3 | 0 |

When a vertical polarizer is placed after each slit (Q3), there will be no horizontally polarized photons that reach the screen. For the vertically polarized photons that reach the screen, WPI is not known and therefore these photons will show an interference pattern at the screen. Since the same number $N$ of photons reach the screen, this interference pattern is no different from the pattern observed when no polarizers are placed after either slit. As shown in Table IX, the most common incorrect answer for both the undergraduate and graduate students is that there will be no interference. A common incorrect reasoning, especially before students worked on the DSE QuILT, is that in this situation, WPI will be known for all photons.

If a vertical polarizer is placed after one slit (say the top slit) and a horizontal polarizer is placed after the other slit (bottom slit) as in Q4, then WPI is known for all photons because a horizontally polarized photon detected at the screen must have gone through the bottom slit and a vertically polarized photon detected at the screen must have gone through the top slit. On this question, the most common incorrect answer was that a full interference pattern should form. Students who provided responses of this type may have had difficulty recognizing that the polarizers provide WPI for all photons, or may believe that even though WPI is known for all photons, an interference pattern is still observed. For example, one graduate student recognized that WPI can be obtained both for a vertically and a horizontally polarized photon detected at the screen, and concluded that neither horizontally nor vertically polarized photons interfere with themselves. However, she thought that they can interfere with each other and said: "I don't know... would they [photons coming from one slit] be able to interfere with the ones [photons] coming from the other slit...?"

When probed further, she said "If it [photon] can only go through one slit or the other it can't interfere with itself, but once it goes through it, there would still be wave propagation [...] would it [a vertically polarized photon] be able to interfere with the horizontally polarized photons or not... I don't know."

When the interviewer asked, "So what you're saying is that a single photon can only go through one slit or the other but you're not sure if that implies that there's no interference because that photon might interfere with another photon that's coming through the other slit, is that right?", she responded, "Yeah."

*Difficulties with quantum eraser*

The last situation (vertical polarizer after one slit, horizontal polarizer after the other, 45° polarizer in front of the screen) is a quantum eraser because the last polarizer erases WPI that could be obtained due to the effect of the other two polarizers. Table XI shows that the most common incorrect answer for both undergraduate and graduate students is that there will be no interference in this situation. Many students who provided these types of responses ignored the third polarizer. For example, one student stated "I don't think interference is possible because you are still identifying the path of one side of photons as different from the other." Another student stated, "See no interference since one is horizontally and the other vertically polarized." These types of reasoning indicate that students essentially ignored the effect of the third polarizer, which erases WPI. As further evidence of students recognizing the similarity between the MZI and

TABLE XI.   Percentages of undergraduate (UG) and graduate students (GS) with different answers to Q5 (full interference, partial interference, no interference, other, and no response/"I don't know"), including percentages of students who mention MZI or quantum eraser when responding to Q5. Bold italic indicates correct response.

|  | Full interference | Partial interference | No interference | Other | No response/ don't know | Mention MZI or quantum eraser |
|---|---|---|---|---|---|---|
| UG-before QuILT | *43* | 2 | 17 | 2 | 36 | 24 |
| UG-after QuILT | *86* | 5 | 2 | 0 | 7 | 66 |
| GS-before QuILT | *52* | 2 | 12 | 12 | 22 | 5 |
| GS-after QuILT | *76* | 10 | 12 | 2 | 0 | 27 |

DSE, in particular, with regards to using WPI reasoning, for this question, many students, especially in the DSE post-test, specifically mentioned the similarity to the MZI, wrote down "quantum eraser" or reasoned in a manner which could have been learned only in the context of the MZI (e.g., the third polarizer erases the WPI obtained from the other two polarizers) even though such things were not mentioned in the DSE QuILT.

---